# Charge ordering transition in GdBaCo$_2$O$_5$: evidence of reentrant behavior

M. Allieta[1*], M. Scavini[1,2*], L. Lopresti[1,2,3], M. Coduri[1], L. Loconte[1], S. Cappelli[1], C. Oliva[1], P. Ghigna[4], P. Pattison[5,6], V. Scagnoli[7,8]

[1]Università degli studi di Milano, Dipartimento di Chimica;

[2]CNR-ISTM, and INSTM, Unit Milan, Via C. Golgi 19, I-20133 Milan, Italy;

[3]Centre for Materials Crystallography, Århus University, Langelandsgade 140, DK-8000 Århus C., Denmark;

[4]Università degli Studi di Pavia, Dipartimento di Chimica, Viale Taramelli 16, 27100 Pavia, Italy;

[5]Swiss-Norwegian Beamlines, European Synchrotron Radiation Facility, 6 rue Jules Horowitz, BP 220, 38043 Grenoble Cedex 9, France

[6]Laboratory of Crystallography, BSP, Ecole polytechnique fédérale de Lausanne, CH-1015 Lausanne, Switzerland

[7]European Synchrotron Radiation Facility, 6 rue Jules Horowitz, BP 220, 38043 Grenoble Cedex 9, France;

[8]Paul Scherrer Institut, CH-5232 Villigen PSI, Switzerland.

**Abstract**

We present a detailed study on the charge ordering (CO) transition in GdBaCo$_2$O$_{5.0}$ system by combining high resolution synchrotron powder/single crystal diffraction with electron paramagnetic resonance (EPR) experiments as a function of temperature. We found a second order structural phase transition at $T_{CO}$=247 K (*Pmmm* to *Pmma*) associated with the onset of long range CO. At $T_{min} \approx 1.2 T_{CO}$, the EPR linewidth rapidly broadens providing evidence of spin fluctuations due to magnetic interactions between Gd$^{3+}$ ions and antiferromagnetic couplings of Co$^{2+}$/Co$^{3+}$ sublattices. This likely indicates that, analogously to manganites, the long-range antiferromagnetic order in GdBaCo$_2$O$_{5.0}$ sets in at $\approx T_{CO}$. Pair distribution function (PDF) analysis of diffraction data revealed signatures of structural inhomogeneities at low temperature. By comparing the average and local bond valences, we found that above $T_{CO}$ the local structure is consistent with a fully random occupation of Co$^{2+}$ and Co$^{3+}$ in a 1:1 ratio and with a complete charge ordering below $T_{CO}$. Below $T \approx 100$ K the charge localization is partially melted at the local scale, suggesting a reentrant behavior of CO. This result is supported by the weakening of superstructure reflections and the temperature evolution of EPR linewidth that is consistent with paramagnetic (PM) reentrant behavior reported in the GdBaCo$_2$O$_{5.5}$ parent compound.

*Corresponding Authors: e-mail: mattia.allieta@gmail.com, marco.scavini@unimi.it

**INTRODUCTION**

Charge ordering (CO) observed in mixed valence perovskites is an intriguing phenomenon which has attracted a lot of interest in the last two decades.[1-6] The stabilization of CO depends on the commensurability of the charge carrier density with respect to the lattice periodicity and is driven by the competition of the inter-site Coulomb and kinetic energies of electrons. The onset of CO phase triggered, e.g., by temperature ($T$), gives rise to an abrupt change in the transport and/or magnetic properties reflecting an intricate interplay between charge, spin, orbital and lattice degrees of freedom.[1-6]

One of the most remarkable examples of perovskitic oxides showing CO is the colossal magnetoresistive (CMR) $R_{0.5}A_{0.5}MnO_3$ (R: trivalent ion, A: Ca, Sr) manganites.[1,3] In these systems the charge carriers localization due to the occurrence of $Mn^{3+}/Mn^{4+}$ CO pattern is accompanied by a sudden increase of the electrical resistivity and by the onset of antiferromagnetic (AFM) phase at $T_N \leq T_{CO}$.[1]

Similar effects have been observed for the layered perovskite-like compounds $Ln$BaCo$_2$O$_{5+\delta}$ ($Ln$: Lanthanide or Y).[5-7] These systems display a larger CMR effect than Mn-based perovskites and a variety of magnetic and transport properties, depending on oxygen concentration $\delta$.[7] In particular, $\delta$ affects the mixed valence state of the cobalt ions, resulting in different $Co^{2+}/Co^{3+}$ and $Co^{4+}/Co^{3+}$ ratios with cobalt species stabilizing in several spin states.[5-9] In the case of $Ln$BaCo$_2$O$_{5.0}$ system ($Ln$= Y[5], Tb[6], Dy[6], Ho[6]), where a $Co^{2+}/Co^{3+}$ ratio equal to 1, neutron powder diffraction (NPD) shows the insurgence of both charge and magnetic ordering associated with structural phase transitions. In the high-$T$ paramagnetic (PM) state, these compounds display a tetragonal structure $P4/mmm$ and with $a \times a \times 2a$ cell metric, where $a$ is the primitive cubic perovskite lattice parameter (Fig 1 (*a*)).[5,6,8] At $T_N \approx 350$ K, these systems undergo a magnetic transition leading to $G$-type AFM phase, associated to an orthorhombic-distorted lattice with $Pmmm$ and the same $a \times a \times 2a$ cell metric (Fig 1 (*b*)). Despite the symmetry lowering, a unique Co site in square pyramidal environment is still present in this phase.[5,6,8] Below $T \approx 210$ K, the NPD measurements suggested the occurrence of another structural phase transition leading to a change of both cell metric and crystal symmetry. In this new low-$T$ phase with *Pmma* and $a \times 2a \times 2a$ cell metric[5,6] (Fig 1 (*c*)), two distinct Co sites are present and the transition can be interpreted in terms of CO.[5,6] The AFM structure in the charge ordered state is qualitatively explained by Goodenough-Kanamori (GK) rules for superexchange and $\approx 1$ $\mu_B$ difference between the refined magnetic moments of the non-equivalent Co sites at saturation is fully consistent with a complete $Co^{2+}/Co^{3+}$ separation.[6] However, the bond valence sums (BVS) calculated for the Co-O distances are far from the ideal +2 and +3 values, suggesting that only partial charge redistribution arises at $T_{CO}$. In addition, the occurrence of AFM ordering at higher $T$ than for long range CO ($T_N > T_{CO}$) is opposite of what is observed in Mn-based compounds. This is the signature of a more complex scenario,[5,6,8] indicating that magnetic, electronic and structural properties of $Ln$BaCo$_2$O$_{5.0}$ systems still remain to be reconciled in a self-consistent picture.

In GdBaCo$_2$O$_{5.0}$ (GBCO), the occurrence of a CO transition below $T \approx 250$ K has been suggested by heat capacity[10] and internal friction[11] measurements. However, the existence of a CO-driven structural phase transition in this compound still remains unproven to date.[10,11] In a previous work,[8] we assigned the correct space group of GBCO at room temperature and we carefully studied the *P4/mmm* to *Pmmm* phase transition across $T_N \approx 350$ K. Here, we focus on the CO transition by combining X-ray powder diffraction (XRPD), synchrotron single crystal X-ray diffraction (SCD) and EPR spectroscopy on GBCO. These techniques are sensitive to either to CO or to the magnetic degree of freedom allowing us to explore the transition both from the structural and spin relaxation point of view.

As a result, we deduced the presence of a second order structural phase transition associated with CO at $T$=247K. By cooling down to $T \approx 100$ K, we observed a local structural response consistent with a reentrant behavior of CO. The EPR measurements suggest that the origin of CO as well as reentrant behavior is driven by an intimate interplay of electron localization and magnetic spin (dis)ordering.

**EXPERIMENTAL**

The same single crystal and powdered GBCO samples employed in our previous study[8] were used for the present diffraction measurements with synchrotron X-rays. SCD experiments were performed at the BM1A station of the Swiss-Norwegian beamline of the European Synchrotron Radiation Facility (ESRF, Grenoble, France). Data acquisitions were recorded by double-crystal Si(111) monochromated X-ray radiation ($\lambda$ = 0.70826(2) Å) on a Kuma KM6-CH six-circle kappa goniometer equipped with an Oxford Diffraction CCD area detector and an Oxford Cryosystems $N_2$ gas blower. The sample was mounted on the top of a glass capillary fibre and SCD data collections were performed at 9 different temperatures between 100 K and 380 K by three $\omega$-scans (132, 136 and 216 deg wide) with $\Delta\omega$ = 1 deg steps at fixed $2\vartheta$, $\kappa$ and $\varphi$ axes (in degrees: 30, 30, 0; 30, –45, 0; 0, 0, 180). Overall, this scheme provided, on average, a ≈ 97% complete dataset within the Cu-sphere resolution ($\sin\vartheta/\lambda$ = 0.652 Å$^{-1}$). Moreover, the temperature evolution of the intensity of the Pmma (116) reflection, extinct in the high-$T$ phases, was monitored by repeated $\varphi$ scans.

CrysAlysPro[12] was employed to perform the data collection and reduction procedures. The unit cell parameters at each temperature (see Table S1 in the SI[17]) were determined from least-square fitting of the orientation matrix against the observed peak positions of, on average, ≈800 intense reflections. Empirical absorption correction based on multiple measures of equivalent reflections was applied to all the data sets according to the SCALE3 ABSPACK algorithm[12] The crystal structure at each temperature was refined with SHELX.[13] Assessment of the sample quality can be found in the SI[17].

XRPD experiments were performed at the ID31 beamline of the ESRF by selecting $\lambda$= 0.39620(5)Å. Twelve patterns in the 0≤$2\vartheta$≤50° range data were collected for 1 hour counting time between 5K and 298K. The sample was cooled down to 80 K by using a $N_2$ gas blower (Oxford Cryosystems), whereas a liquid-helium-cooled cryostat was employed to achieve the lowest temperature. Moreover, some patterns were collected at $\lambda$= 0.35422(5) Å in the $80 \leq T \leq 400$ K range. In particular, at selected temperatures ($T=80$, $T=180$K and $T=298$ K) several scans were summed up for 7 hours counting time ($Q_{max} \approx 27$ Å$^{-1}$) to achieve statistical significance for Pair distribution function (PDF) analysis.

Data were analyzed using the Rietveld method as implemented in the GSAS software suite.[14] Absorption correction was performed through the Lobanov empirical formula implemented for the Debye-Scherrer geometry. Line profiles were fitted using a modified pseudo-Voigt function accounting for asymmetry correction. In the last refinement cycles, scale factor(s), cell parameters, positional coordinates and isotropic thermal parameters were allowed to vary as well as background and line profile parameters.

EPR measurements were performed at a Bruker ELEXSYS spectrometer equipped with an ER4102ST standard rectangular cavity at *X* band (9.4 GHz) frequency every 5 K in the temperature range 115-450 K. The derivative d*P*/d*H* of power P absorbed was recorded as a function of the static magnetic field *H*.

## RESULTS

### I. SCD and XRPD across the CO transition

To probe the occurrence of low *T* structural phase transitions associated with possible CO effects, the reciprocal lattice as determined from SCD data was carefully screened to detect possible superstructure reflections. Actually, some commensurate superstructure diffraction spots with *h*/2 indices clearly appeared below 250 K, implying the low temperature doubling of the cell axis. Conversely, no recognizable superstructure peaks were detected above $T \approx 250$ K.

SCD structural models as a function of *T* were obtained within the spherical atom approximation,[16] with the thermal motion of heavy metal atoms modelled as anisotropic. The *Pmmm* room-temperature structure was employed as a suitable starting point for refining the positional and thermal motion parameters down to 250 K. Below $T = 250$ K, we adopted the structural *Pmma* model by Fauth et al.[6] as a starting point for the least-square refinement procedure.[6] Table S1 of the SI[17] summarizes the final statistics of the structural refinements, together with relevant details of the SCD experiments.

As reported in a similar structural study,[15] the transition from *Pmmm* to *Pmma* with $a \times 2a \times 2a$ unit cell can be followed by the temperature dependence of (*hkl*) reflections with $h = 2n+1$ and $l \neq 0$. Figure 2 (*a*) shows the temperature dependence of the intensity of the *Pmma* (116) superstructure reflection, that is extinct in the *Pmmm* phase. On the basis of these results, $T_{CO}=247$ K can be assumed as a reasonable estimate for the CO transition temperature, as at this temperature the (116) superstructure reflection begins to be clearly significant ($\approx$ 2 esd's) with respect to the background. According to the critical equation $I(T)=I(0)(T_c-T/T_c)^\beta$ with $T_c = T_{CO} = 247$ K, the least-square estimate of the critical exponent $\beta$ came out as large as 0.483(5) (inset of Fig.2(*a*)), quite close to the ideal value for a second-order transition ($\beta = 0.5$). It is worth noting that, for $T < \approx 120$ K, the temperature dependence of (116) shows a slightly positive slope.

Clear signature of the CO transition was also found by analyzing the XRPD patterns through, e.g., the appearance of the *Pmma* (102) reflection across $T_{CO}$ (Fig.2(*b*), inset). An accurate investigation of XRPD patterns does not reveal further peaks splitting as well as superstructure reflections corresponding to a modulation of $a \times 2a \times 2a$ periodicity down to 5 K.

The structural parameters obtained from SCD data were used as starting values to perform Rietveld refinements against XRPD data as a function of temperature. Figure 3 shows an example of the Rietveld refinement performed at 5 K. From the quality of the fit it can be seen that XRPD patterns in the CO state can be well reproduced down to the lowest temperature by the *Pmma* superstructure. In Table S2 of the SI[17], selected structural data and agreement factors obtained for all patterns collected at different temperatures are listed. In *Pmmm* and *Pmma* structural models, we constrained isotropic thermal parameters related to oxygen positions to be the same.

In Fig. 4, we present the refined lattice parameters and cell volume as a function of *T* obtained by SCD and XRPD. A reduced orthorhombic cell metric was used for comparison purposed between the *Pmmm* and *Pmma* structures. As previously reported, the structure is tetragonal at $T>T_N$ and with decreasing $T$ the orthorhombic distortion turns out to be appreciable in terms of the lattice metrics.[8] In particular, below $T_N$ the distortion becomes more evident and the difference between *a* and *b* cell edges increases toward the CO transition (see Fig.4) The *a*-axis suddenly increases while the *b*-axis steeply decreases at $T < T_{CO}$ and this anisotropic thermal expansion of the unit cell parameters is maintained down to $T \approx 100$ K. Below this temperature, both the axes reach a plateau by approaching constant values down to $T = 5$ K. This complex behaviour suggests an intricate interplay between CO and the atomic interactions on the *ab*-plane of GBCO.

It worth noting that, as shown in Fig.4 (a), (c), there are some differences between the results obtained from the SCD and XRPD techniques. By approaching the $T_{CO}$ from above, *a* and *b* axes obtained from the two techniques are in fairly good agreement up to 298 K, but for $T < T_{CO}$ the anisotropic thermal expansion of the unit cell parameters seems to be different. In particular, in SCD results also the *b* lattice parameter exhibits a small lengthening across transition, resulting in an overall increase of the unit cell volume at $T_{CO}$ (Fig.4 (c), full circles). This behavior was not detected by XRPD that, on the contrary, found only a weak variation of the slope of the temperature dependence of *V* across $T_{CO}$. However, looking at the general trends of the structural parameters across the CO transition (Fig. 4), we can say that – at least qualitatively – SCD and XRPD gave similar results. In the following discussion, we will focus on the Co valence states and their interplay with the corresponding average Co coordination geometries.

Selected Co-O distances ($d_{Co-O}$) as a function of $T$ are reported in Fig.5(*a*), (*b*). The presence of two distinct Co crystallographic sites in the *Pmma* CO phase gives rise to an anisotropic thermal expansion of $d_{Co-O}$. In particular, by decreasing $T$ below $T_{CO}$, the $d_{Co2-O1/O2}$ expands while $d_{Co1-O1/O2}$ suddenly shrinks. Both distances approach constant values down to 5K. The same behavior is shown by the mean Co-O distances ($<d_{Co-O}>$) related to the Co2 and Co1 sites (Fig.5(c)). The evolution of $d_{Co-O}$ across $T_{CO}$ provides evidence of a smaller volume of the $Co1O_5$ with respect to the Co2O$_5$ square pyramid. According to tabulated Shannon ionic radii,[18] this is consistent with Co1 and Co2 sites occupied by Co$^{3+}$ and Co$^{2+}$, respectively.

To evaluate ionic charges of Co from the experimental $d_{Co-O}$, we calculated BVS by using the tabulated parameters.[19] The results for all the cations of GBCO are shown in Fig. 6. BVS calculations for Gd and Ba cations gave reliable results yielding to valences of ≈2.9 and ≈2.1, respectively. No change in BVSs was observed above and below the CO transition, as shown in the inset of Fig. 6. Above $T_{CO}$, BVS related to Co provides evidence that the sites are occupied by cations in mixed valence +2.3, a value slightly lower than that expected for a 1:1 ratio of Co$^{2+}$/Co$^{3+}$. Below $T_{CO}$, BVS found valences as large as ≈ 2.6 for the Co1 site and ≈ 2.0 for the Co2 site down to 5 K. According to Fauth et al.,[6] the *Pmma* structure can be described by an ordered alternation of Co$^{2+}$O$_5$ and Co$^{3+}$O$_5$ pyramids stacked along the *a* and *c* axes, while both Co$^{2+}$O$_5$ and Co$^{3+}$O$_5$ pyramids are allowed to run along the [010] direction (Fig. 1).

It should be noted that the estimated valence of the Co$^{2+}$/Co$^{3}$ and Co1 sites are markedly lower than +2.5 and +3 expected values, respectively. Such deviations from formal reference values are rather common in perovskite compounds[20] and the observed valence depletion can be possibly due to a charge disorder affecting the Co sites. Moreover, this behavior is consistent with data reported for Ho and Tb parent compounds,[6] for which the calculated BVS suggest partial charge redistribution at both Co1 and Co2 sites.

**II. PDF across the charge ordering transition**

PDF analysis of the XRPD data collected at $T$ = 80 K, 180 K, 298 K was carried out using the formalism of $G(r)$ functions. $G(r)$ is obtained via the sine Fourier Transform (FT) according to :

$$G(r) = \frac{2}{\pi} \int_0^{Q_{max}} Q[S(Q)-1]\sin(Qr) dQ \tag{1}$$

where $S(Q)$ is the total scattering structure function, $Q=4\pi \sin\theta/\lambda$ and $r$ is defined in the space of interatomic distances. $S(Q)$ is calculated from the experimental scattering intensity $I^{coh}(Q)$ containing both Bragg and diffuse scattering contributions. To consistently evaluate $I^{coh}(Q)$, the raw diffracted intensity profile $I(Q)$ collected at each temperature was corrected for background scattering, attenuation in the sample, multiple and Compton scattering. The reduction process was done using the PDFGetX2 software[21] and full-structure profile refinements were carried out on the observed $G(r)$ using the PDFgui program.[22] The program assesses the degree of accuracy of the refinement by the agreement factor $R_w$ defined as:

$$R_W = \left[\frac{\sum w_i(G_i^{exp} - G_i^{calc})^2}{\sum w_i(G_i^{exp})^2}\right]^{1/2} \qquad (2)$$

Figure 7 (a) shows the experimental PDF curves obtained in the 1.3Å≤r≤10Å range. Each positive peak in $G(r)$ function is proportional to the probability of finding two atoms separated by a distance $r$ averaged over all pairs of atoms in the sample. The main temperature induced fluctuations in G(r) curves are given by the peak sharpening observed upon cooling. This is consistent with Debye law and therefore it cannot be considered as a response of the PDF to CO transition.

The experimental profiles were fitted in the 1.8 ≤ r ≤ 30 Å range using the same structural models employed for interpreting the SCD and XRPD patterns. In general, they gave a good description of the PDF in a wide range of $r$, as testified by the good $R_w$ values obtained: 0.082 ($T$=80 K), 0.083 ($T$=180 K) and 0.079 ($T$=298 K). Focusing on the very short $r$ range (1.8 ≤ r ≤ 2.6 Å), Gd-O distances (at ≈ 2.4 Å) have weak $T$ dependence in agreement with the reciprocal space analysis (see Fig.S1 in the SI). On the other hand, the short range Co-O next neighbor distances (in the 1.9 ≤ r ≤ 2.1 Å range) show interesting features as a function of $T$. At 298 K PDF displays just a single peak near 2 Å besides the termination ripple at $r ≈ 2.15$ Å (Fig. 8(*a*)). At 180 K a bimodal distribution of the same PDF peak is found in agreement with the CO *Pmma* structure (Fig.8(*b*)). Indeed, below $T_{CO}$ the originally single <Co-O> PDF peak splits into different bond length distributions (see Fig. 5 and the marked peak in Fig.8(*b*)). The integrated area ratio of these two new peaks is ≈ ½, suggesting a similar multiplicity of the atom pair *correlation*, consistent with different $Co^{3+}/Co^{2+}$ oxygen bond distances. PDF analysis confirms that two distinct sites for $Co^{3+}$ and $Co^{2+}$ ions should be present at $T$ = 180 K. Conversely, discrepancies between average and local structure are found at 80K. The Rietveld analysis clearly indicates that the structure keeps two independent Co sites at this temperature but the experimental PDF at $T$=80K (Fig. 8(*c*)) exhibits a single-modal Co-O distance distribution. Moreover, the width of this peak is clearly increased in respect to higher $T$ PDFs and is comparable with the overall width of the corresponding double peak at 180 K. This implies that the local average charge distribution of $Co^{2+}/Co^{3+}$ ions differs from the long-range and that some kind of disorder should be present.

In order to get more quantitative information about the point charge distribution we estimated the *local* BVS and compared it with the *long* range BVS obtained from the mean <Co-O> distances reported in Fig. 5(*c*). In these calculations, we constrained the Co-O distances in the pyramid to be the same. The inset of Fig.8 shows the local, average BVS together with expected Co charges for a full disordered (solid line) and full ordered (dashed line) phases. First, local BVS at $T$ = 298K is in very good agreement with a complete random occupation of $Co^{2+}$ and $Co^{3+}$ in a 1:1 ratio, whereas at $T$ = 180 K it is consistent with a complete CO. On the other hand, at $T$ = 80 K local BVS becomes very similar to the long-range BVS above $T_{CO}$. However, the increased width of Co-O peak suggests the contribution of several locally different Co-O distances due to a partial melting of charge localization at this temperature. This can be explained in terms of a weaker localization of the charge carriers, resulting in a smaller charge separation among symmetry-independent Co sites or, equivalently, in a reduction of the CO degree between Co atoms in the real space.

Accordingly, the intensity of the superstructure (102) reflection, that is greater than zero just in the *Pmma* phase, rapidly increases below ≈ 250K reaching a maximum at $T$ ≈ 100 K (Fig. 2(b)). Below this temperature, its intensity begins to slightly decrease. At $T$ ≈ 5 K it roughly shows the same value as that measured at ≈ 200 K, which is ≈ 60% of its maximum value. The observed weakening of the (102) intensity is then consistent with a reduction of the CO state strength and agrees with the PDF outcome at $T$=80K.

**III EPR across the charge ordering transition**

Figure 9 shows the EPR spectra as a function of temperature for GBCO sample. As found in $GdBaCo_2O_{5+\delta}$ compounds with $\delta \geq 0.5$, the signal consists of a single broad resonance line originated from the shift and/or broadening of the $Gd^{3+}$ resonance, that in turn is caused by exchange interaction between localized 4$f$ electron spin and the spins of Co atoms.[9] Looking at the temperature dependence of the EPR spectra, it is clear that the signals markedly change across the structural transitions. In general, the available analytical functions well describe the shape of EPR spectra at each temperature but the fits performed gave meaningless parameters. We decided to extract the peak-to-peak linewidth ($\Delta H_{pp}$) as a direct observation of EPR data as depicted in the inset of Fig.9. Moreover, since in our spectra the baseline is not well defined, we cannot directly determine the peak position. Hence, we will consider only the $\Delta H_{pp}$ parameter throughout.

The temperature dependence of $\Delta H_{pp}$ is shown in Fig.10. In high-$T$ PM regime regime (370 K ≤ $T$ ≤ 450 K), $\Delta H_{pp}$ decreases linearly by following the Korringa-type relation[23] $\Delta H_{pp} = \Delta H_{pp}^0 + bT$ with $\Delta H_{pp}^0$= 2858(8) G and $b$= 2.27(2) G/K. By further cooling, $\Delta H_{pp}$ shows a departure from the linear narrowing. On approaching $T_{CO}$ upon cooling, $\Delta H_{pp}$ shows a weak drop at $T \approx$ 365 K close to $T_N$ and it goes through a minimum at $T_{min} \approx$ 300 K. Below $T_{min.}$ the $\Delta H_{pp}$ has a complex thermal behavior. It displays a fast broadening reaching a maximum value at $T \approx T_{CO}$ and then rapidly decreases below $T_{CO}$. Interestingly, below 160K the $\Delta H_{pp}$ seems to follow the linear temperature dependence observed at high $T$ in the PM phase.

**DISCUSSION**

GBCO undergoes two structural phase transitions upon cooling. In our previous investigation[8] we found the first crystallographic transformation associated with a second-order phase transition. In particular, at $T_N \approx 350K$ the lsymmetry lowers from tetragonal (*P4/mmm*) to orthorhombic (*Pmmm*), inducing the loss of the $C_4$ axis along **c**. The breaking of the tetragonal symmetry perturbs the coordination environment around Gd and Co, and one half of oxygen atoms on the $CoO_2$ plane become symmetry-independent at $T < T_N$.

Below $T_{CO}$=247 K, a second structural phase transition takes place associated with the CO of the $Co^{2+}/Co^{3+}$ ions. The space group changes from *Pmmm* to *Pmma*, with the simultaneous doubling of *a* axis. The analysis of the temperature dependence of the intensity of the *Pmma* (116) superstructure reflection across this transition showed a critical exponent close to the ideal value for a second-order transition. However, Fauth *et al.*[6] studied the CO structural phase transition in the Tb, Dy and Ho parent compounds[6] and some differences have to be noted. First, the transition temperature is significantly higher for the $GdBaCo_2O_5$ system ($T_{CO}$ = 247 K vs $T_{CO}$ = 205-215 K) suggesting that $T_{CO}$ increases with the size of the rare earth ion. It should be noted that such an increase is at least three times higher on passing from Tb to Gd ($\approx$ 30 K) than from Ho to Tb ($\approx$ 10 K). Secondly, Fauth *et al.*[6] interpreted the CO phase transition as first order on the basis of their DSC measurements on the $HoBaCo_2O_5$ sample. All of this evidence suggest that the effect of $Gd^{3+}$ on the transport and magnetic properties of the GBCO structure is critical and it implies a significant increase of the range of existence of the low-*T* CO *Pmma* phase.

By further cooling, we found the signatures of a new structural anomaly. The depletion of the *Pmma* (102) superstructure reflection below $T \approx 100$ K indicates that the CO state emerges at $T_{CO}$ but weakens below $\approx$ 100 K. Accordingly, at $T = 80$ K the outcomes of our PDF analysis suggest that GBCO seems to recover a charge-disordered phase at the local scale. Due to the large width of the peak, the distance distribution of the Co-O first coordination shell is still consistent with the *Pmma* structural model. However, the corresponding local BVS value suggests a partial melting of the charge localization. This structural difference between the local and the long range scales is well known to occur in many strongly correlated perovskite-like oxides showing properties such as magnetodieletric[24] or superconductivity[25].

Reentrant CO behavior has been reported on $LaSr_2Mn_2O_7$ layered perovskite and its origin has been linked to the occurrence of a ferromagnetic-metallic state upon cooling.[2] As for the present case, possible reasons of this behavior can be investigated by considering the temperature evolution of spin relaxation as probed by EPR. Different relaxation mechanisms are known to result in a linear temperature dependence of $\Delta H_{pp}$ above the CO transition.[23] The decreasing $\Delta H_{pp}$ on cooling can be mediated by the exchange scattering between the local moments (4f $Gd^{3+}$ localized states) and itinerant charge carriers of Co ions. In the absence of bottleneck and dynamic effects, the $b$ in Korringa-type relation is given by:[23]

$$b = \frac{\pi k_B}{g\mu_B} N(E_F)^2 J^2 \qquad (3)$$

where the term $J_{fs}$ is the exchange coupling constant between the conduction electrons and localized spin and $N(E_F)$ stands for the density states at the Fermi energy.

From the linear fit of our EPR data in $360 \leq T \leq 450$ K range we found a low value of $b \approx 2.7$ G/K which suggests also a low $N(E_F)$ value. This is consistent with the fairly low resistivity reported for $GdBaCo_2O_{5+\delta}$ ($\delta \to 0$) systems ($\approx 0.01$ $\Omega$ cm at $T = 370$ K[7]), which implies a non-negligible delocalization for conduction electron in GBCO. To account for the Korringa-type behavior in charge disordered PM phase, we consider that the electron delocalization of conduction electrons can be due to the motion of an extra electron from $e_g$ level of $Co^{2+}$ ion in high spin (HS) configurations to an empty $e_g$ orbital of $Co^{3+}$ ion in intermediate spin (IS) state. Note that these simple models imply a double exchange (DE) interactions between Co ions in different spin states, and they have been widely used to explain the insulator-to-metal transition observed in GdBaCo$_2$O$_{5+\delta}$ with $\delta \approx 0.5$.[9] The charge disordered GBCO phase is characterized by a random distribution of Co$^{2+}$/Co$^{3+}$ and $e_g$ electron hopping from an HSCo$^{2+}$ to next ISCo$^{3+}$ can give rise to electron delocalization in all the spatial directions. These isotropic DE paths between two pyramidal sites, defined as A and A can be sketched as:

$$t_{2g}^5 e_g^2\text{-HSCo}^{2+}{}_A + t_{2g}^5 e_g^1\text{-ISCo}^{3+}{}_B \rightarrow t_{2g}^5 e_g^1\text{-ISCo}^{3+}{}_A + t_{2g}^5 e_g^2\text{-HSCo}^{2+}{}_B \qquad (4)$$

It is worth to note that these exchange channels are reminiscent of the well known DE between Mn$^{3+}$ and Mn$^{4+}$ invoked to correlate magnetic and electronic properties in mixed valence CO manganites.[1]

While the PM contribution of Gd$^{3+}$ dominates the magnetization of GBCO, the temperature dependence of magnetization of YBaCo$_2$O$_5$ – the only parent compound with non magnetic ions – shows two drops at $T_N$ and $T_{CO}$.[5,7] $\Delta H_{pp}$ data observed on GBCO show a weak upturn at $T \approx 365$ K, which possibly mimics the drop observed on YBaCo$_2$O$_5$ magnetization around $T_N$.[5,7] This temperature above $T_N$ perfectly matches with the onset of the tetragonal (200) peak broadening precursor of the $P4/mmm$ to $Pmmm$ transition.[8] This provides a clear evidence which links the onset of spin ordering to the structural phase transitions in GBCO.[8]

Basically, the emergence of AFM can be discussed by assuming that the structural transition is accompanied by a crossover from DE to super-exchange Co-O-Co interactions. According to GK rules[6] the relevant super-exchange couplings which hold for the AFM interactions in the $Pmmm$ phase are between HS-Co$^{2+}$-ISCo$^{3+}$ along the [001] direction and HSCo$^{2+}$-HSCo$^{2+}$, IS-Co$^{3+}$-ISCo$^{3+}$ along the [010] direction.[6] However, these magnetic interactions can be considered dominant only if a preliminary Co$^{3+}$/Co$^{2+}$ ordering is assumed at $T_N$. Actually, incomplete or short range charge ordering has been invoked to explain the resistivity measurements in the $T_{CO} \leq T \leq T_N$ range of LnBaCo$_2$O$_5$ parent compound.[6] However, we did not observe any signature of the CO phase in the short range PDF of GBCO at RT.

AFM interactions between Co ions and partial conduction electron delocalization can generate short-range AFM correlation between $Gd^{3+}$ through the Ruderman–Kittel–Kasuya–Yosida (RKKY) interaction. In the $T_{CO} \leq T \leq T_N$ range, the conduction electrons are mobile to some extent and they can transfer the Co AFM spin fluctuations to the Gd sites *via* the RKKY mechanism. This gives rise to an effective alternating internal magnetic field which relaxes the Gd spins. In principle, upon cooling one could expect the strengthening of AFM interaction between Co ions and an increase of the spin relaxation rate of $Gd^{3+}$, which produces the broadening of the EPR signal. This is exactly what is observed in our EPR data below $T_{min} \approx 300$ K (see Fig.10). The EPR line broadening while decreasing $T$ has been interpreted as a precursor behavior of AFM transitions.[26] According to available theories,[26] the critical increase of $\Delta H_{pp}$ due to AFM spin fluctuations can be described by the relation $\Delta H_{pp}=a[(T-T_N)/T_N]^{-m}$ where $a$ and $m$ are constants. In our data, $\Delta H_{pp}$ starts to increase at $T_{min.} \sim 1.2 T_{CO}$ and rapidly broadens as $T_{CO}$ is approached from above. In the inset of Fig.10 the log-log plot of $\Delta H_{pp}$ as a function of $[(T-T_{CO})/T_{CO}]$ is shown. A linear correlation is evident just above $T_{CO}$, analogue to that expected near $T_N$ for AFM spin fluctuations.[26] We determined $m = -0.12$ in agreement with that found in other layered compounds where the $Gd^{3+}$ spin dynamics is affected by 3d metal spins fluctuations.[26,27] The CO transition in GBCO seems to be related to spin fluctuations involving $Gd^{3+}$ ions. This implies that $T_N \approx 250$ K $\approx T_{CO}$ should be set as the actual Néel temperature. When $T$ is further lowered, long-range AFM order is driven by a long range CO. On the other hand, we speculate that the transition at 350K is more related to the onset of incomplete CO giving rise to short range AFM interaction, as also previously suggested.[8] This is consistent with the magnetization of $YBaCo_2O_5$ which shows a more marked drop at $T_{CO}$ than $T_N$.[5,7]

The emergence of long-range CO is therefore associated with a strong localization of the conduction electrons that weakens the RKKY coupling between Co and Gd below $T_{CO}$. Since the super-exchange interactions are stronger and not frustrated within the 3D CO network,[6] the overall AFM coupling increases with CO distortion. The internal field owing to spin fluctuations gradually disappears and the EPR signal is dominated by the PM contribution of $Gd^{3+}$ spins, which narrows $\Delta H_{pp}$ on cooling. Below 160 K, $\Delta H_{pp}$ seems to follow the same linear temperature dependence observed at high-$T$. Looking at our data in the low-$T$ region, the temperature dependence of $\Delta H_{pp}$ is indeed similar to that observed in the PM phase. Reentrant PM behavior has been also observed at $T$ = 75 K in $GdBaCo_2O_{5.5}$ and the transition has been associated to a spin state transition (SST) from IS to low spin (LS) $t_{2g}^6 e_g^0$ state at $Co^{3+}$ site.[28] SST involving the same $Co^{3+}$ spin states was also reported for $LaCoO_3$ at $T \approx 100$ K.[29] Moreover, the switch from IS to LS of $Co^{3+}$ introduces non-magnetic ions within the super-exchange framework of the CO phase of GBCO. As a result, super-exchange couplings involving $Co^{3+}$ disappears, disrupting the AFM order that gradually weakens on cooling. Since in the CO phase the electron localization is mainly driven by AFM couplings between ordered $Co^{2+}/Co^{3+}$, the emergence of PM phase at low-$T$ well matches the partial melting of charges localization observed from structural analysis.

To explain the interplay between the structural and magnetic ground state evolution as a function of $T$ in GBCO we propose the following mechanism. According to DE between the randomly distributed $HSCo^{2+}$ and $ISCo^{3+}$, the conduction electrons in the high-$T$ PM phase are delocalized to some extent. On going through $T_N \approx 350$ K, the $Co^{2+}/Co^{3+}$ ions start to partially order, leading to the coexistence of both localized and delocalized charge carriers. Because of incomplete CO, the crossover from DE to super-exchange interactions gives rise to RKKY couplings between short-range AFM and $Gd^{3+}$ spins. At $T_{min} \approx 1.2T_{CO}$ the emergence of AFM spin fluctuations are a precursor effect of the AFM transition associated with the onset of CO. Once the long range CO is initiated the density of delocalized conduction electron rapidly decreases. As a result, $\Delta H_{pp}$ decreases on cooling and below $T \approx 160$ K it gradually approaches the linear $T$-dependence observed in the high-$T$ PM phase. Below this point, we observed the mismatch between the local and the average BVS and the decrease of superstructure reflection intensity suggesting a reentrant behaviour of CO. These latter observations, together with $T_N \approx T_{CO}$, leads us to believe that CO layered cobaltite shows strong similarity with manganites.[1-4]

**CONCLUSION**

In the present study we have clearly revealed the intimate interplay of electron localization and magnetic spin ordering in GBCO. The second order *Pmmm* to *Pmma* structural phase transition associated with CO at $T_{CO}$=247 K was directly observed by SCD and XRPD techniques. In particular, the emerging of the CO phase is supported by (i) the behavior of the superstructure reflections, (ii) the temperature dependence of the Co-O distances and (iii) the computed BVS. EPR measurements gave important insights into the CO transition holding the emergence of spin fluctuations at $T_{min} \approx 1.2\ T_{CO}$. As reported for other layered compounds,[26,27] the spin fluctuations can be considered a precursor effect of AFM transition associated to the onset of CO in GBCO system. Hence, in agreement with the $YBaCo_2O_5$ magnetization curve, which shows a more marked drop at $T_{CO}$ than $T_N$,[5,7] we suggest that the long range AFM order takes place at $\approx T_{CO}$ in GBCO. PDF analysis of diffraction data shows new structural features. At 298K and 180K the local structure is consistent with a fully random occupation of $Co^{2+}$ and $Co^{3+}$ in a 1:1 ratio and with a complete CO. On the other hand, at *T*=80K the local range PDF seems to be consistent with a melting of charges localization. In turn, this suggests that a reentrant CO transition occurs in GBCO. This behaviour, analogous to that found in CMR manganites,[2] is supported by the temperature dependence of $\Delta H_{pp}$ below 160K and is consistent with the PM reentrant behavior found for the $GdBaCo_2O_{5.5}$ parent compound.

**Acknowledgements**

The authors gratefully acknowledge the European Synchrotron Radiation Facility for provision of beam time and Dr. Adrian Hill for assistance in using the ID31 beamline. The authors would like to thank Dr D. Chernyshov for useful discussions. The Danish National Research Foundation through the Center for Materials Crystallography (CMC) has also been also very much appreciated.

**FIGURES**

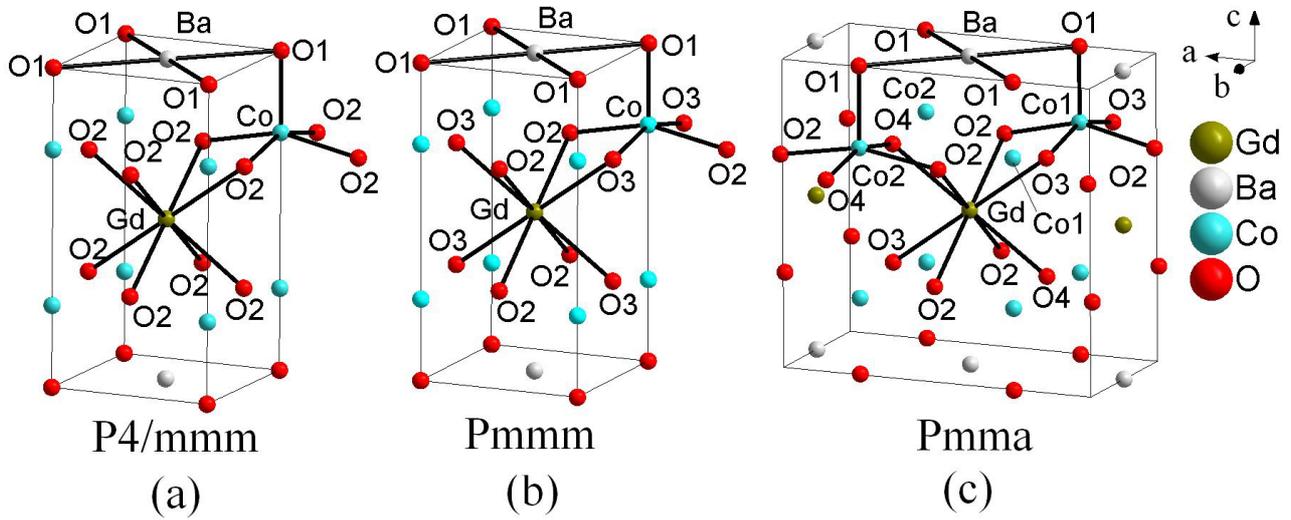

FIG.1 (Colour online) Packing and atom numbering schemes for each of the three phases of GdBaCo$_2$O$_5$ this work: *P*4/*mmm* (a), *Pmmm* (b) and *Pmma* (c). For each symmetry-independent metal ion, its next-neighbour Oxygen coordination environment is highlighted by black bonds.

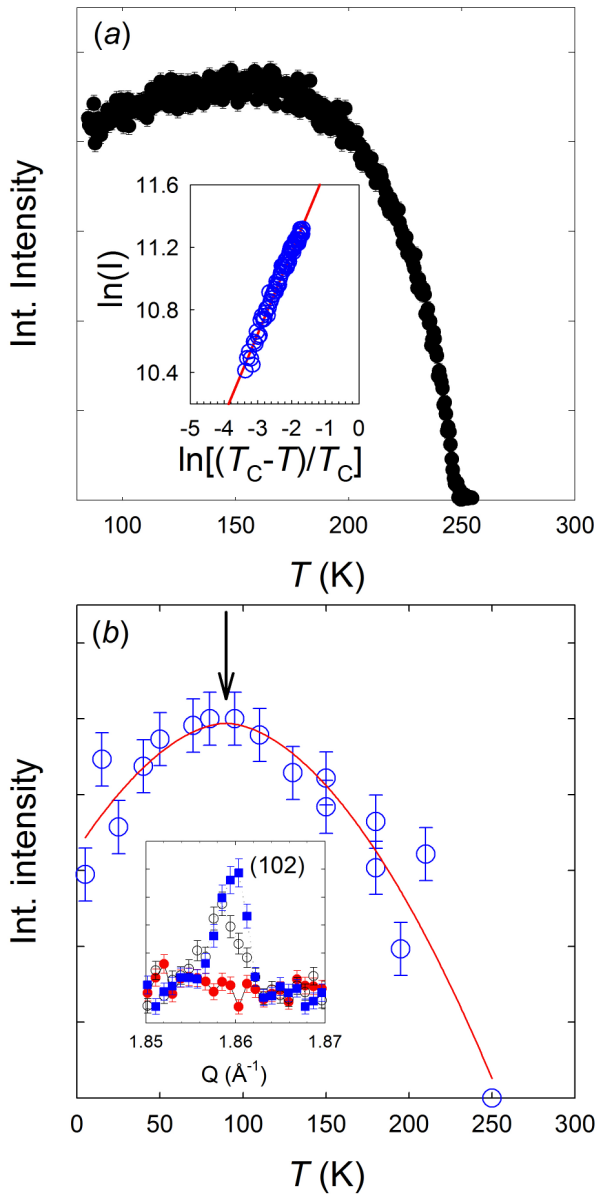

FIG.2 (Colour online) (a) Temperature dependence of intensity of the *Pmma* (116) superstructure reflection (circles). Inset: ln-ln plot of the intensity versus $(T_c-T)/T_c$ . (b) The temperature evolution of *Pmma* (102) reflection as collected at ID31 is reported as example (empty circles). The arrow indicates the onset of the intensity weakening. The inset shows XRPD profiles related to (102) reflection across the CO transition. Note that the intensity of superstructure reflections is much higher than the error bars given by the background.

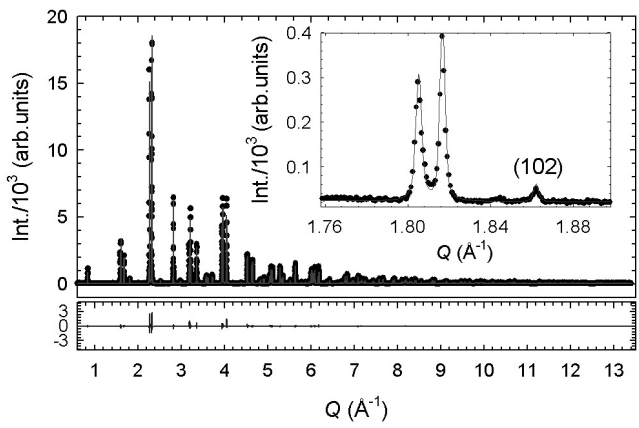

FIG. 3 Measured (dots), calculated (line) powder diffraction patterns and residuals (bottom line) for GBCO at $T$=5K and at $\lambda$= 0.39620(5) Å. The inset shows a magnified view of the low angle diffraction peaks where the CO superlattice reflection is clearly shown.

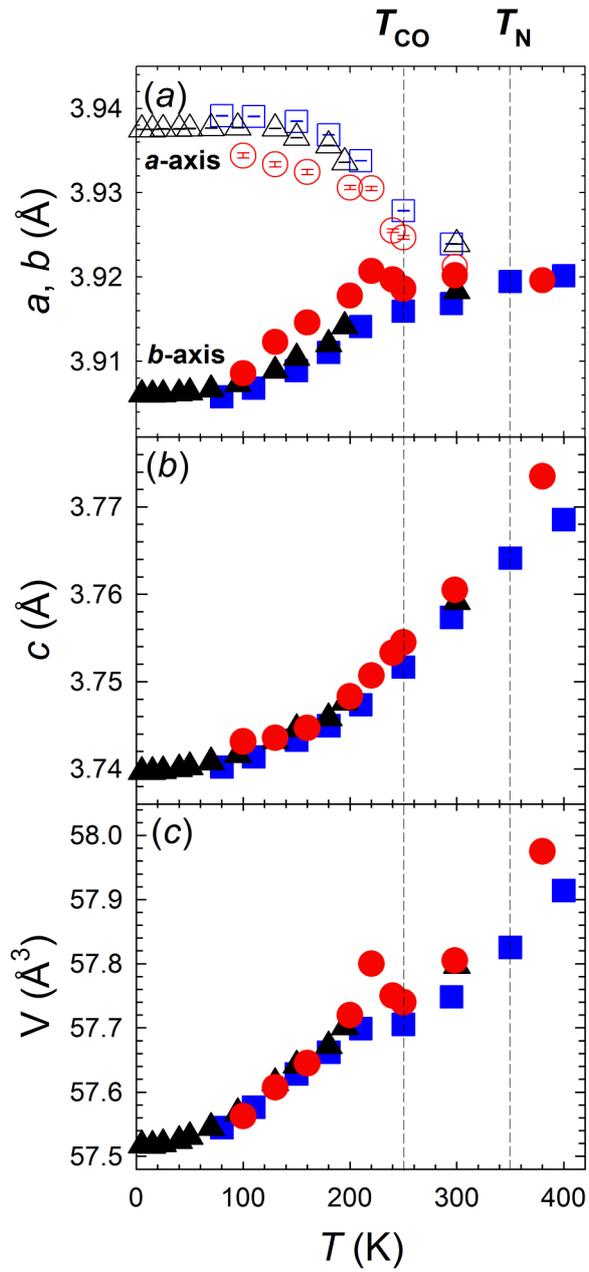

FIG.4 Temperature dependence of the (a) *a, b* axes (b) *c*-axis and (c) unit cell volume. Full and empty circles are data from SCD; full and empty squares and triangles are data from XRPD.

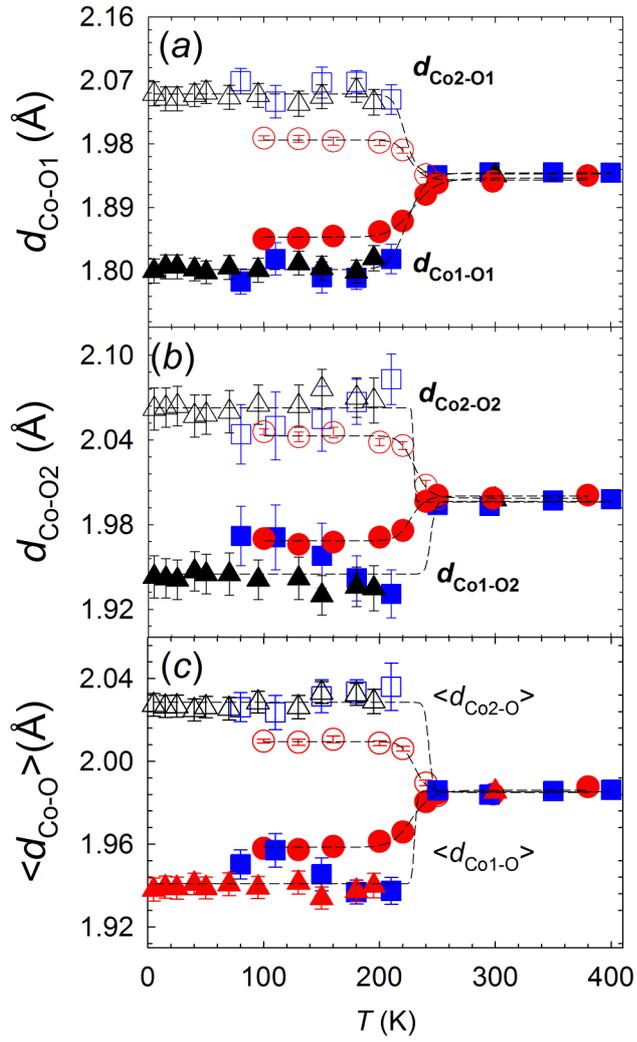

FIG.5. Temperature dependence of (a), (b) selected Co-O and (c) average <Co-O> distances. Full and empty circles are data from SCD; full and empty squares and triangles are data from XRPD. Dashed lines are guide to the eye.

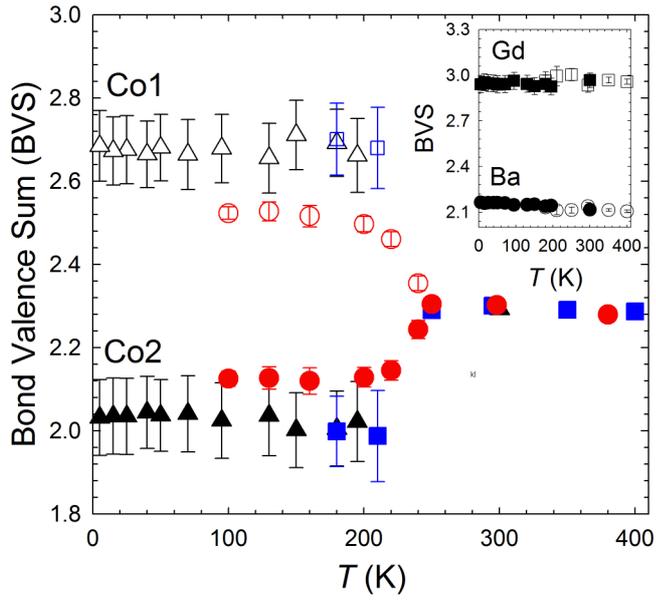

FIG.6 Temperature dependence of calculated Bond Valence Sum (BVS) for Co in GBCO. Full and empty circles are data from SCD; full and empty squares and triangles are data from XRPD. In the inset the BVS related to Gd and Ba are also shown.

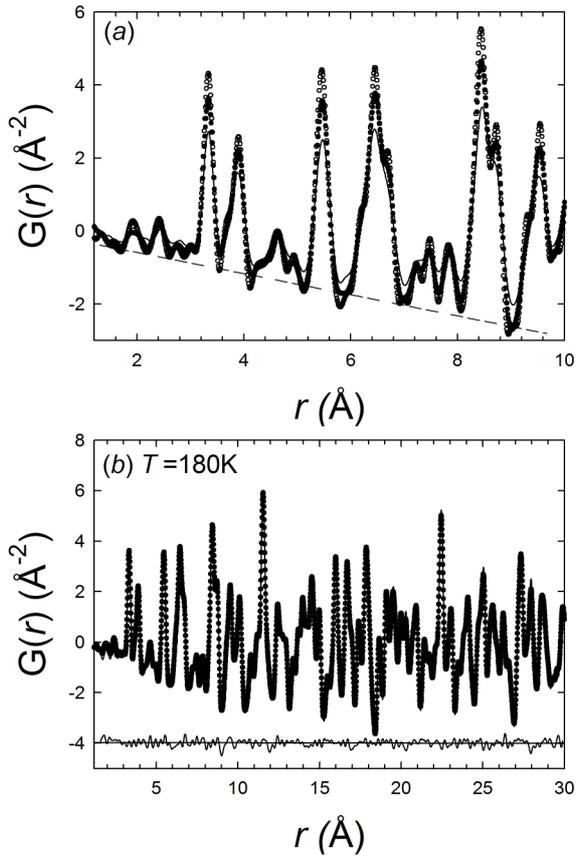

FIG.7 (*a*) PDFs obtained from GBCO sample at *T*=298K (continuous line), *T*=180K (filled circles), *T*=80K (empty circles) within $r = 10$ Å. The dashed line is the baseline and corresponds to $-4\pi r\rho_0$.

(*b*) Observed (dots) and calculated (continuous line) PDF obtained at 180K within $r = 30$ Å. The residual plot is shown at the bottom of the figure.

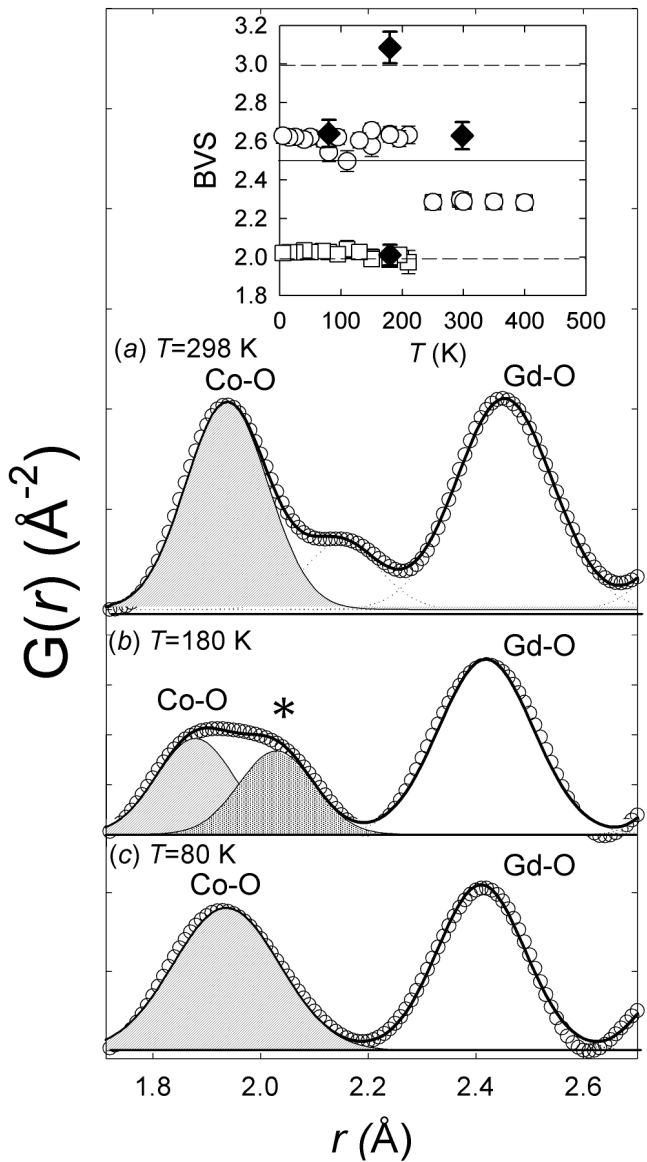

FIG.8 PDFs at short range: empty circles are the experimental curves and continuous lines are the fits shown the Co-O and Gd-O components of the PDF peaks. The '*' marked feature indicates the appearance of a bimodal <Co-O> bond length distribution (see text). The inset shows the local (diamods) and long-range (empty circles and squares) Bond Valence Sums (BVS) for Co as a function of $T$. Straight lines mark the expected BVS values for random and fully ordered $Co^{2+}/Co^{3+}$ charge distributions (see text).

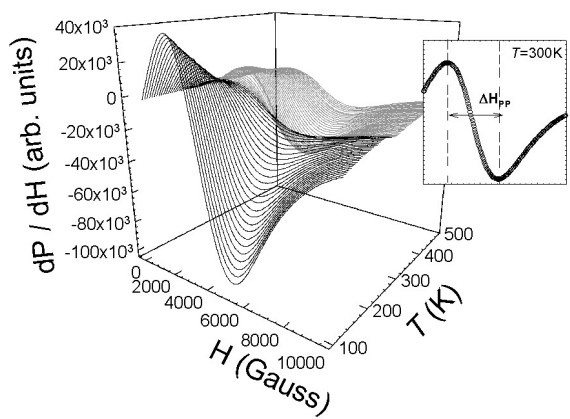

FIG.9 EPR spectra of the GBCO sample as a function of temperature. In the inset an example of observed EPR is shown.

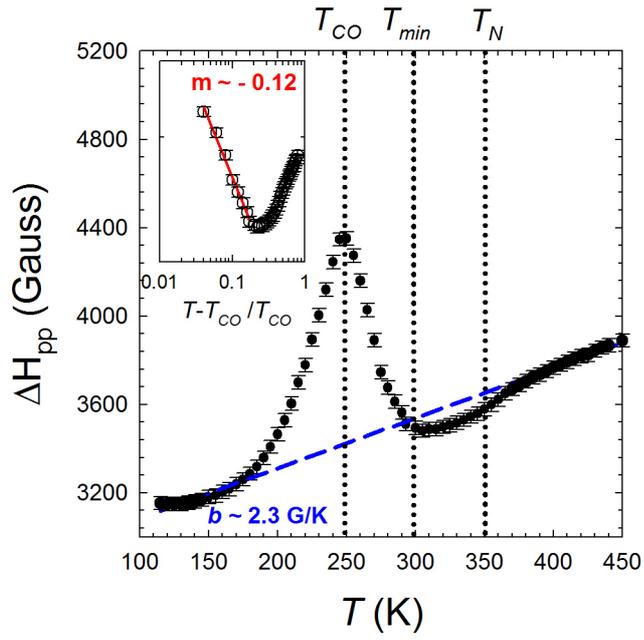

FIG. 10 Temperature evolution of $\Delta H_{pp}$ for GBCO. The inset shows the log-log of the $\Delta H_{pp}$ as a function of $(T - T_{CO}/T_{CO})$.